\documentclass{article}
\usepackage{preamble}
\usepackage{booktabs}
\usetikzlibrary{matrix, positioning, shapes.geometric}
\usepackage{tikz-3dplot}
\usepackage{graphicx}
\usepackage{tikz}
\usepackage{url}
\usepackage[left=2cm, right=2cm]{geometry}
\usepackage{float}
\usepackage{enumitem}
\setlist[enumerate]{itemsep=0em}
\usepackage[numbers,sort&compress]{natbib}

\begin{document}

\title{signDNE: A python package for ariaDNE and its sign-oriented extension}

\author{Felix Risbro Hjerrild$^{1}$, Shan Shan$^{1}$, Doug M Boyer$^{2}$, Ingrid Daubechies$^{3}$ \\
\\
$^{1}$Department of Mathematics and Computer Science, University of Southern Denmark \\
$^{2}$ Department of Evolutionary Anthropology, Duke University \\
$^{3}$ Department of Mathematics, Duke University 
}

\maketitle

\vspace{0.3in}

\section*{Abstract}
\begin{enumerate}
	\item A key challenge in evolutionary biology is to develop robust computational tools that can accurately analyze shape variations across diverse anatomical structures. The Dirichlet Normal Energy (DNE) is a shape complexity metric that addresses this by summarizing the local curvature of surfaces, particularly aiding the analytical studies and providing insights into evolutionary and functional adaptations.
	\item Building on the DNE concept, we introduce a Python-based implementation, designed to compute both the original DNE and a newly developed sign-oriented DNE metric. This Python package includes a user-friendly command line interface (CLI) and built-in visualization tools to facilitate the interpretation of the surface's local curvature properties. The addition of signDNE, which integrates the convexity and concavity of surfaces, enhances the tool's ability to identify fine-scale features across a broad range of biological structures.
	\item We validate the robustness of our method by comparing its performance with standard implementations on a dataset of triangular meshes with varying discrete representations. Additionally, we demonstrate its potential applications through visualization of the local curvature field (i.e., local curvature value over the surface) on various biological specimens, showing how it effectively captures complex biological features.
	\item Our goal is to provide evolutionary biologists with enhanced computational tools and visualization capabilities for studying biological form complexity. In this paper, we offer a brief overview of the Python CLI for ease of use. Alongside the Python implementation, we have also {updated the original ariaDNE MATLAB package to ensure consistent and accurate DNE computation across platforms.}These improvements enhance the tool's flexibility, reduce sensitivity to sampling density and mesh quality, and support a more accurate interpretation of biological surface topography.
\end{enumerate}

\noindent
\textbf{Key words: ariaDNE, signDNE, Dirichlet Normal Energy, Shape complexity metric, Morphology}

\section{Introduction}

Biological shapes and anatomical structures are fundamental to the study of evolutionary biology. Numerical descriptors that quantify the overall geometry of biological forms are essential tools for the modeling, analysis, and understanding of evolutionary processes. These numerical descriptors are especially useful when the shapes of study have complex structures and are highly diverse, making standard landmark-based or correspondence-map-based geometric morphometric methods challenging to apply. One significant example of such numerical descriptors is Dirichlet normal energy (DNE) (\cite{bunn2011comparing}), which is a shape complexity metric that summarises the local curvature over the entire surface. 
DNE was originally proposed as a dental topographic metric and has shown considerable potential in terms of inferring dietary relationships (\cite{winchester2014dental, berthaume2016food, lopez2018dental, de2021widespread, selig2021mammalian, selig2024variation}). More recently, it has been applied to other anatomical structures, aiding in the analysis of functional and evolutionary relationships, reconstructing evolutionary pathways, and identifying adaptive traits (\cite{stamos2020ontogeny, thomas2020physical, chiaradia2023tissue, clear2023baculum, pamfilie2023quantifying, assemat2023shape}). 

Given the increasing diversity and widespread availability of 3D shape data, it is essential to ensure that the algorithmic implementation of shape complexity metrics remains robust against variations amongst the data input and the pre-processing procedures. {The sensitivity of the original DNE implementation to these variations led the development of the MATLAB-based ariaDNE (\cite{shan2019ariadne}), which aimed to provide a more reliable computation of DNE. However, as the demand for accessible and open-source tools has grown, it became clear that further advancements in other computational platforms were needed.} In this paper, we introduce a new Python-based implementation of ariaDNE, which faithfully reproduces the original algorithm while offering a more user-friendly experience. This version features an intuitive command line interface (CLI) and includes built-in visualization capabilities for the DNE field, allowing users to easily capture and interpret local curvature across complex anatomical surfaces. 

Sign-oriented DNE, or signDNE (\cite{pampush2022sign}), is a recent development in MolaR that integrates the sign of local curvature to distinguish between concavity and convexity, adding an important layer of detail to the numerical descriptor. However, this implementation, like the original DNE, is sensitive to variations in the discrete representation of continuous shapes. Typically, continuous surfaces are represented as triangular meshes, where a tile of connected triangular faces is used to approximate the continuous surface geometry. The MolaR implementation, however, is particularly sensitive to differences in mesh resolution, the number of triangular faces, surface smoothing, and small noise during data processing. This sensitivity can compromise the accuracy of the downstream analysis and the subsequent results. In this paper, we address these issues by proposing a new method of computing the signDNE, and we demonstrate the robustness of our method on various mesh perturbations. {Additionally, we have incorporated this new feature into the Python package and updated the original ariaDNE MATLAB implementation, ensuring more accurate and stable computation of signDNE across multiple platforms.} 

\begin{figure}
	\centering
	\begin{tikzpicture}
		\matrix[matrix of nodes,
		column sep=1em,
		nodes={align=center}] (m)
		{
			\node{\includegraphics[width=0.21\linewidth]{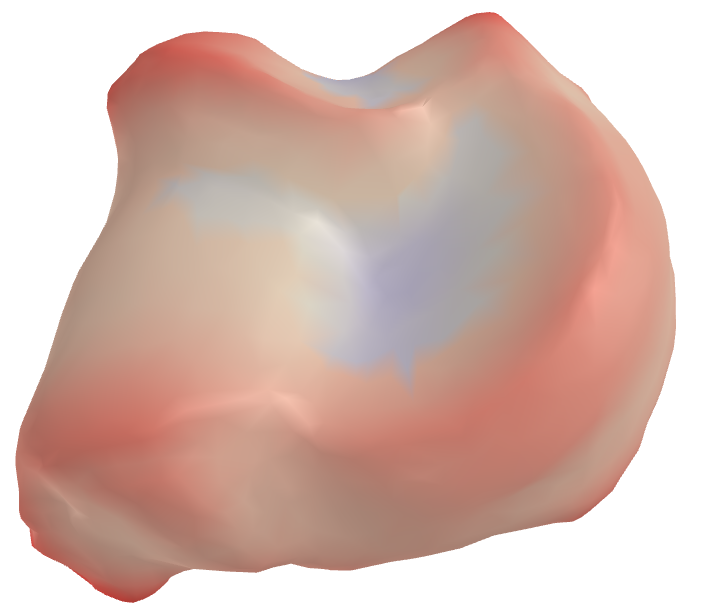}}; &
			\node{\includegraphics[width=0.26\linewidth]{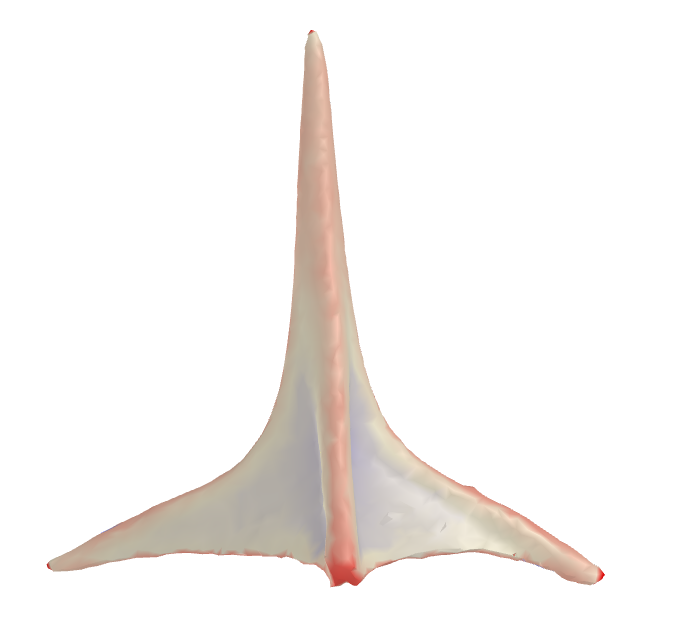}}; &
			\node{\includegraphics[width=0.3\linewidth]{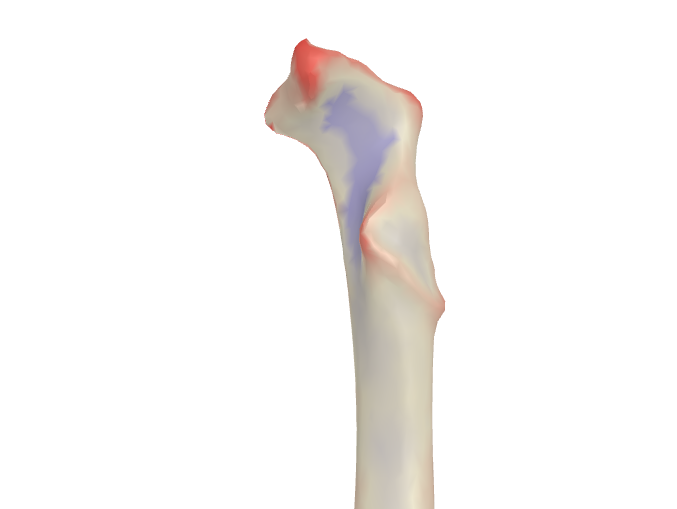}};
			\\
			\node{Mule deer wrist bone \\ (ID: 000122274)}; & \node{Pufferfish spine \\ (ID: 000576115)}; & \node{Greater sage-grouse \\ shoulder bone (ID: 000120565)};
			\\
			\node{\includegraphics[width=0.3\linewidth]{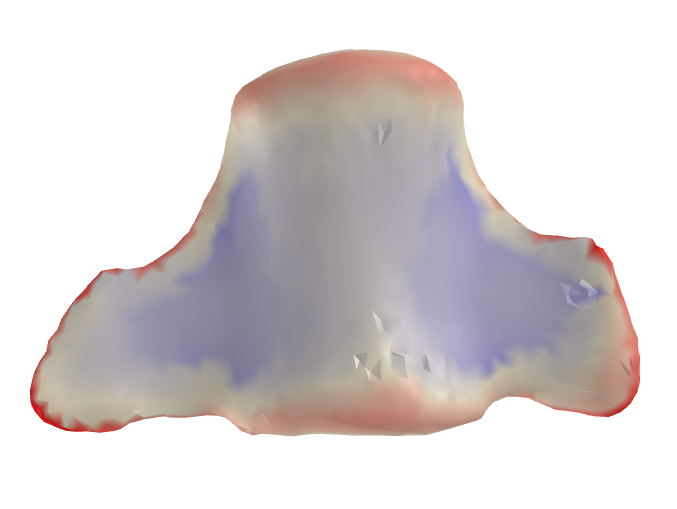}}; & 
			\node{\includegraphics[width=0.23\linewidth]{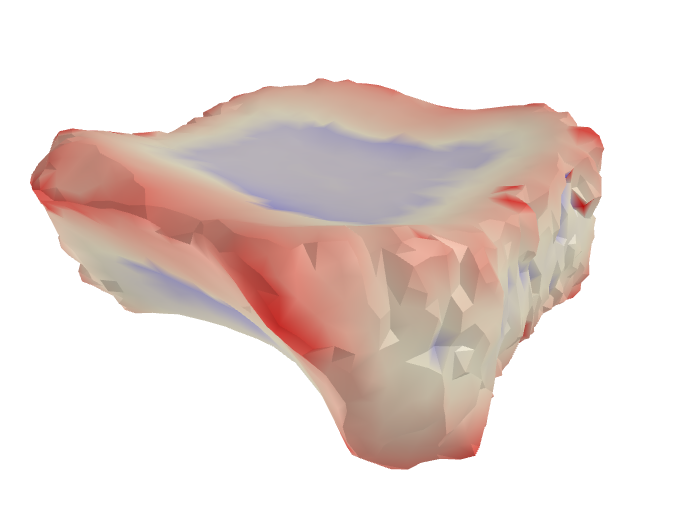}};
			&
			\node{\includegraphics[width=0.19\linewidth]{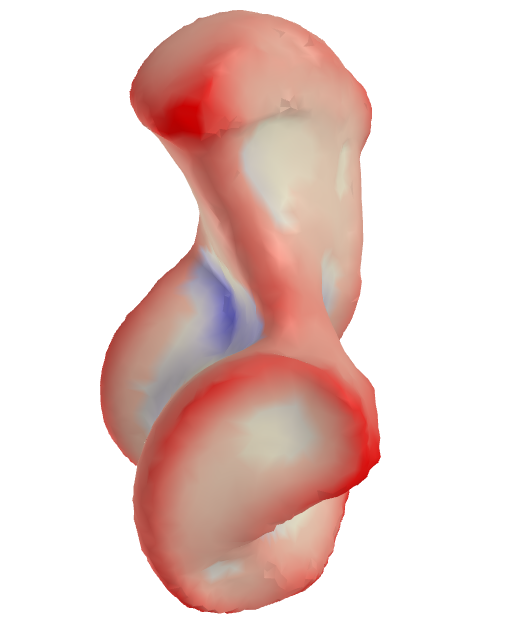}};
			\\
			\node{Muskox tail vertebra \\ (ID: 000121104)}; & \node{Mountain goat chest \\ area bone (ID: 000119433)}; & \node{Fox astragalus \\ (ID: 000458120)};
			\\
		};
	\end{tikzpicture}
	\caption{
		Visualization of local curvature field on various biological specimens. Positive curvature regions are shaded red, and negative curvature regions are shaded blue. The ID's are MorphoSource media identifiers.}
	\label{fig:demovis}
\end{figure}
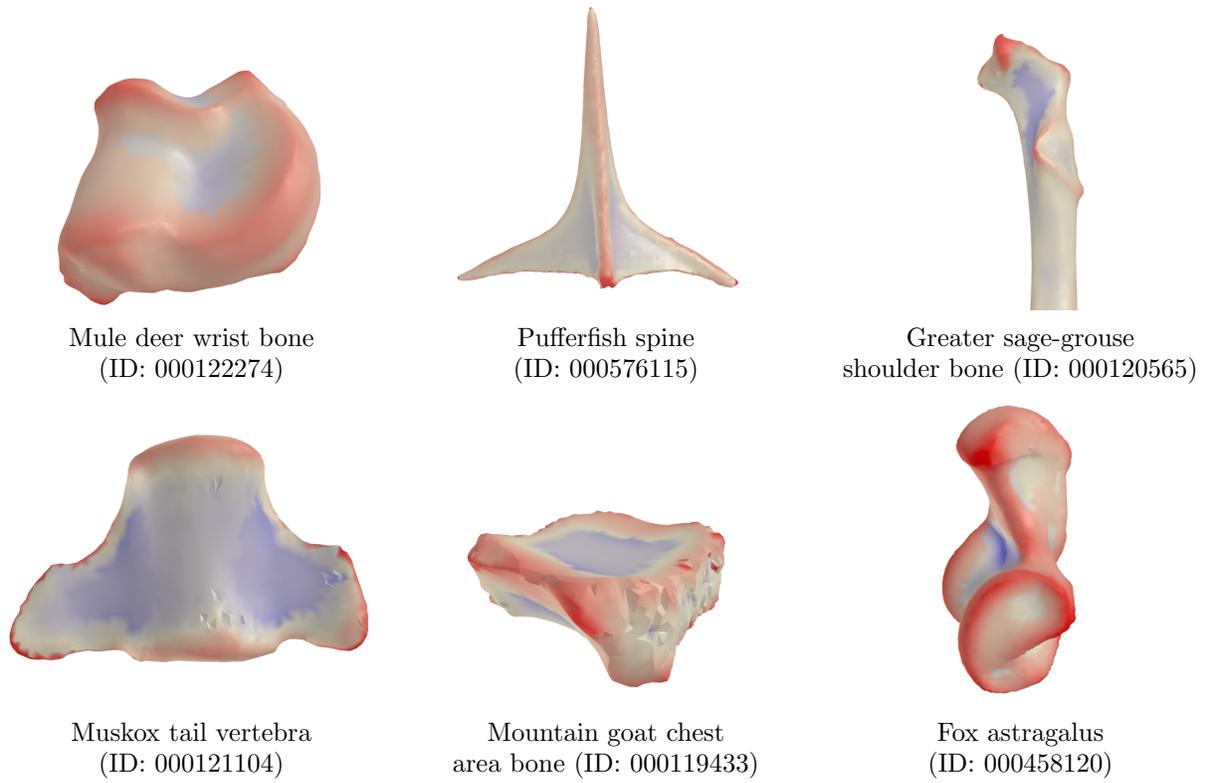

In summary, this paper presents a new Python-based implementation of DNE, which enhances accessibility, usability, and visualization capabilities while faithfully reproducing the original algorithm. Additionally, we improve the computation of signDNE by developing a more robust method that reduces sensitivity to mesh resolution, smoothing, and noise. These advancements provide the scientific community with more reliable and versatile tools for studying biological shape complexity and evolutionary processes. Figure \ref{fig:demovis} illustrates the visualization functionality of the local DNE field on various biological specimens. Figure 2 demonstrates the robustness of our method on various mesh perturbations. 

\begin{figure}
	\centering
	\begin{tikzpicture}
		\matrix[matrix of nodes,
		column sep=0.5em,
		nodes={align=center}] (m)
		{
			\node {}; & 
			\node {Typical}; & 
			\node {2K Tri}; &
			\node {20k Tri}; &
			\node {$10^{-3}$ Noise}; &
			\node {$2 \cdot 10^{-3}$ Noise}; &
			\node {Smooth}; \\
			\node[rotate=90, anchor=center, xshift=0.7cm] {\textbf{signDNE}}; &
			\node {\includegraphics[width=.12\textwidth]{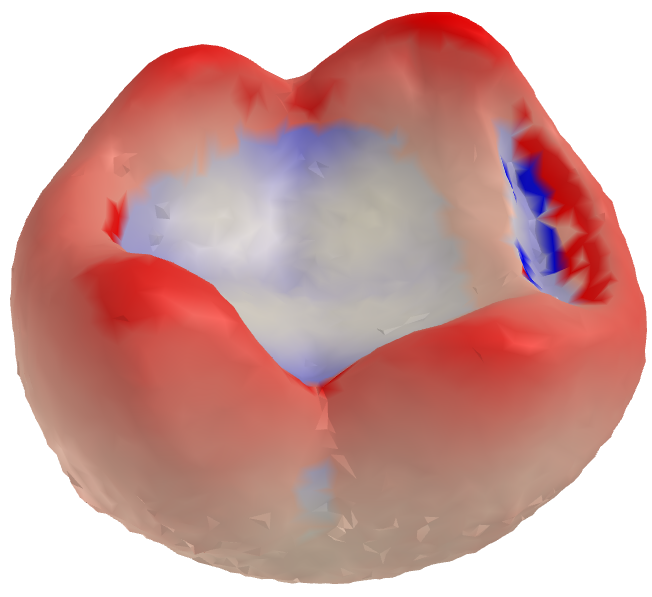}}; &
			\node {\includegraphics[width=.12\textwidth]{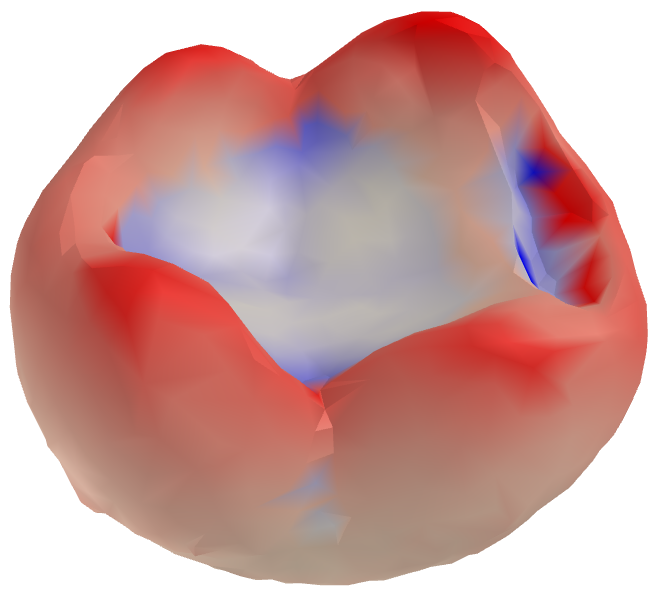}}; &
			\node {\includegraphics[width=.12\textwidth]{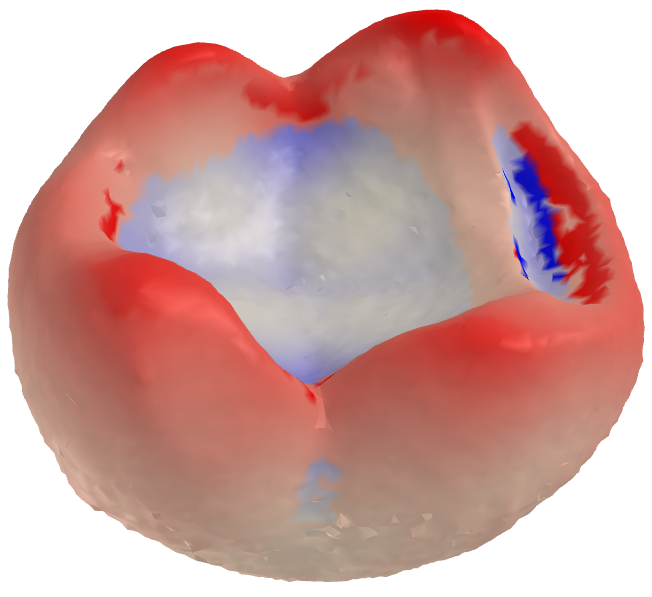}}; &
			\node {\includegraphics[width=.12\textwidth]{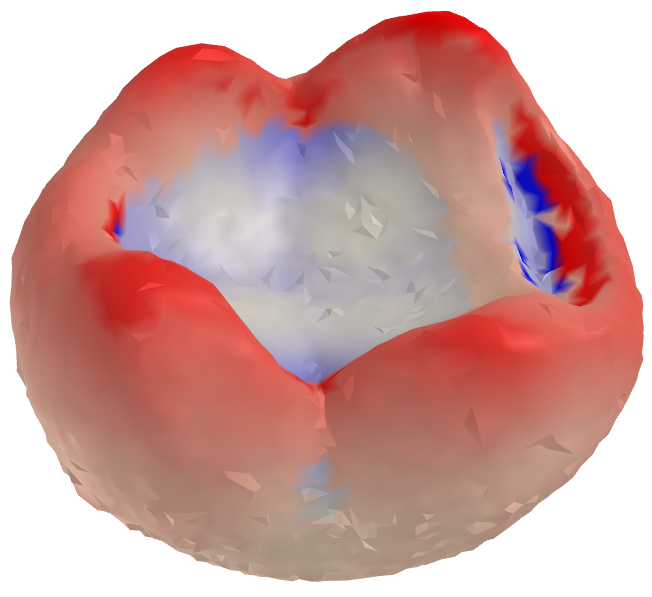}}; &
			\node {\includegraphics[width=.12\textwidth]{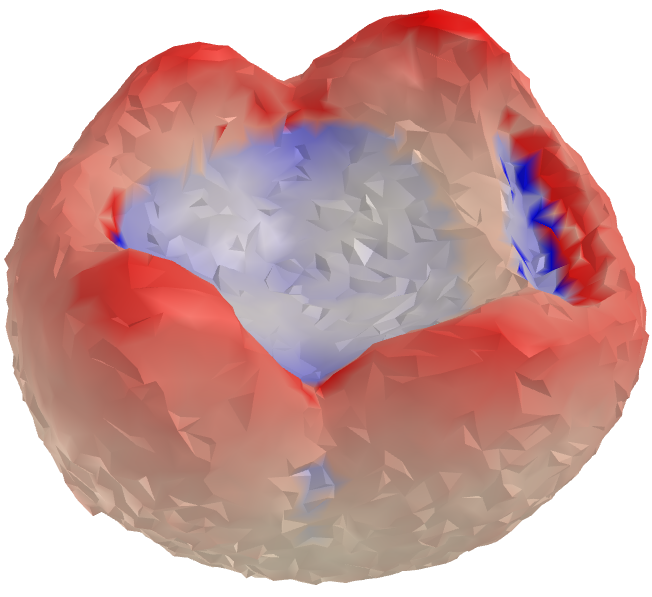}}; &
			\node {\includegraphics[width=.12\textwidth]{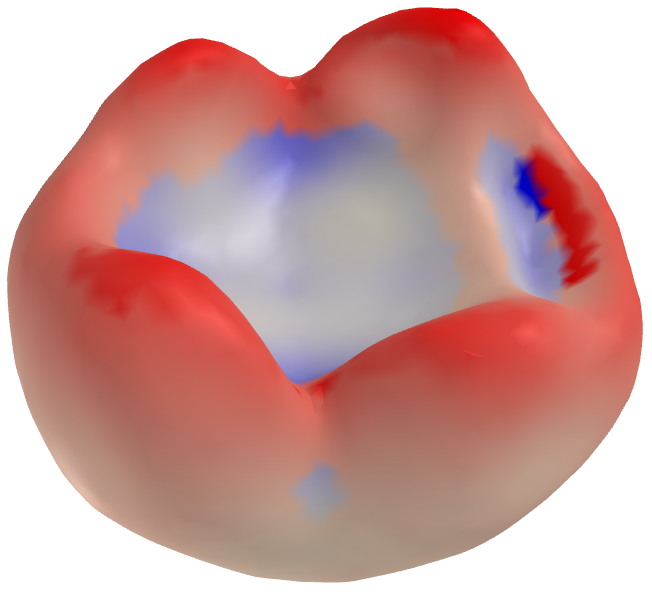}}; \\
		};
	\end{tikzpicture}
	\label{fig:comparison_visual}
	\caption{Visualization of local curvature field on varying triangle count, added noise and smoothing.
	}
\end{figure}
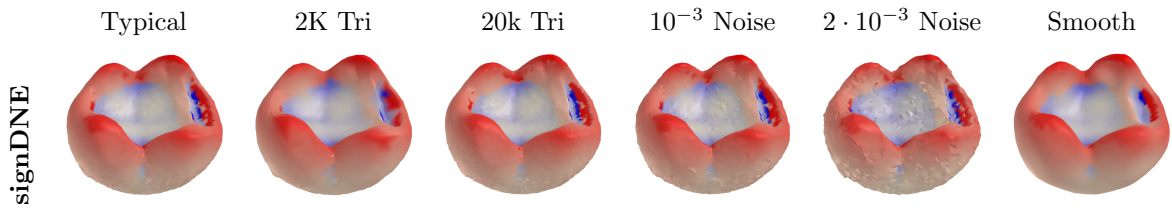

\section{Algorithm description}
DNE is a numerical descriptor defined on a continuous surface, integrating changes in the normal direction at each point of the surface. These changes indicate the extent of local bending. A robust method to approximate the local bending at a given point is to apply principal component analysis (PCA) to a small neighborhood around that point. The plane spanned by the first two principal components approximates the tangent plane of the surface at that point. Local bending is estimated by measuring the deviation of the tangent plane from the surface. Precisely, it is computed by taking the smallest eigenvalue and dividing it by the sum of all eigenvalues. The DNE of the entire surface is then obtained by integrating these values across all points on the surface. The method used in ariaDNE (\cite{shan2019ariadne}) improves this approach in two ways. First, ariaDNE uses weighted PCA, which further increases robustness in low triangle count meshes. The weight function used in ariaDNE is the Gaussian kernel $f(x)=e^{-x^2/\epsilon}$. The $\epsilon$ value, also known as the \textit{bandwidth}, determines the influence of the neighboring points. A greater $\epsilon$ results in increased influence of neighboring points. Second, ariaDNE uses the eigenvalue corresponding to the principle component closest to the normal at each point, instead of the smallest eigenvalue, for computing the local bending. This further ensures the accuracy of the selected normal direction.

The sign-oriented DNE (\cite{pampush2022sign}) is a natural extension of DNE, that distinguishes between the convex and concave regions of a surface by assigning a signed value to each point. Specifically, it assigns a positive sign when the surface bends outward -- like a cusp or a ridge, and a negative sign when the surface bends inward -- like a valley. This assignment can be made by using the sign of the mean curvature, which is mathematically defined as the average of the maximum and minimum principal curvatures at that point. However, conventional mean curvature implementations often lack the robustness necessary for biological shape analysis, particularly when dealing with noisy data or data processed with different protocols.
To address this limitation, we propose a new, more reliable method for determining the sign of the curvature at each point. The steps of our method are as follows:
\begin{enumerate}
	\item \textbf{Compute local bending score and weighted centroid.} For each vertex, compute the Gaussian kernel with a pre-specified bandwidth value $\epsilon$. Using this weighting, calculate the local bending score as in ariaDNE, and then determine the weighted centroid.
	\item \textbf{Ensure watertight mesh.} If the shape has boundaries or holes, we first close them using a hole-filling algorithm to create a watertight mesh.
	\item \textbf{Determine the sign.} Apply ray casting to determine whether the centroid lies inside or outside the watertight version of the mesh. Precisely, shoot a random ray that starts from the identified centroid point, and determine how many times ($n$) the ray passes through the watertight mesh. If $n$ is odd, then the centroid lies inside the watertight mesh, and we assign a negative sign to the vertex; if $n$ is even, then it lies outside the watertight mesh, and we assign a positive sign to the vertex. 
	\item \textbf{Compute signDNE.} The negative component of DNE is computed by summing the local DNE values over the region with negative sign, and likewise, the positive component of DNE is computed by summing the local DNE values over the region 
	with the positive sign. 
\end{enumerate}

\begin{figure}[htb]
	\centering
	\begin{tikzpicture}
		\matrix[matrix of nodes,
		column sep=0.4em,
		nodes={align=center}] (m)
		{
			\node
			{
				\textbf{Step 1:} Compute \\ the local weighted \\ centroid.}; &
			\node{\textbf{Step 2}: Close any \\ holes, to make the \\ mesh watertight.}; &
			\node{\textbf{Step 3:} Determine \\ the curvature sign.\\}; &
			\node{\textbf{Step 4:} Compute \\ the signDNE.\\};
			\\
			\node {\includegraphics[width=.23\textwidth]{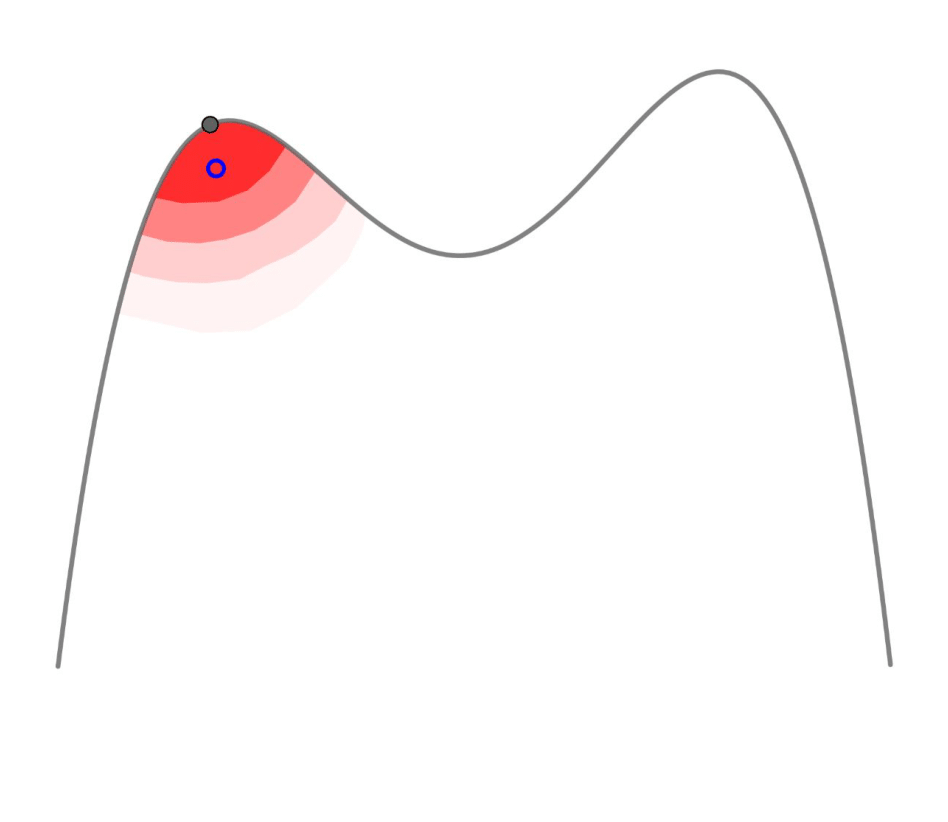}}; &
			\node {\includegraphics[width=.23\textwidth]{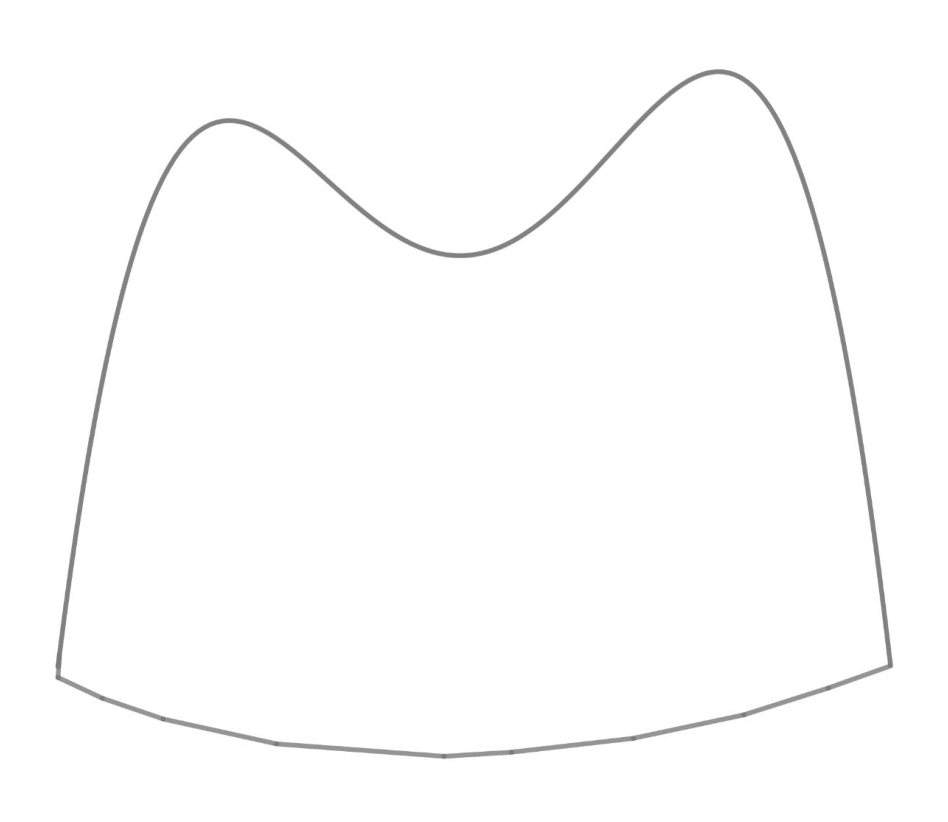}}; &
			\node{\includegraphics[width=.23\textwidth]{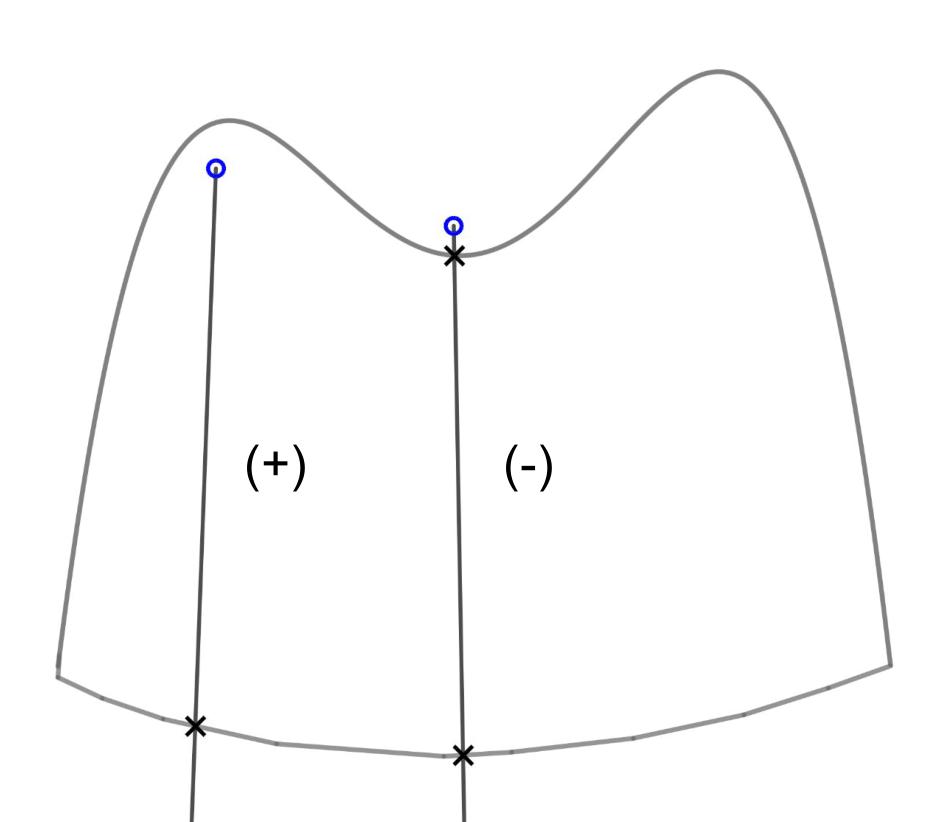}}; &
			\node{\includegraphics[width=.23\textwidth]{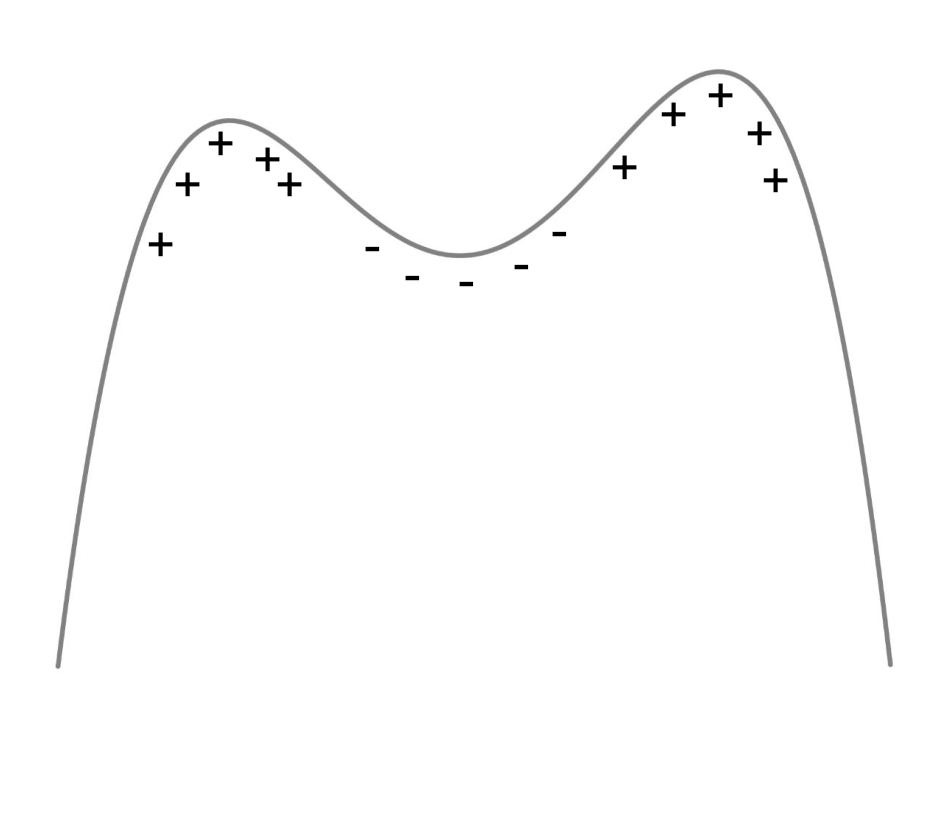}};
			\\
		};
	\end{tikzpicture}
	\caption{Key steps in our proposed signDNE algorithm. The red color graded region represents weighting given by the Gaussian kernel around the current vertex (black dot). The open blue dot(s) represents local centroids, and black crosses is ray intersections with the closed mesh.}
	\label{fig:method}
\end{figure}
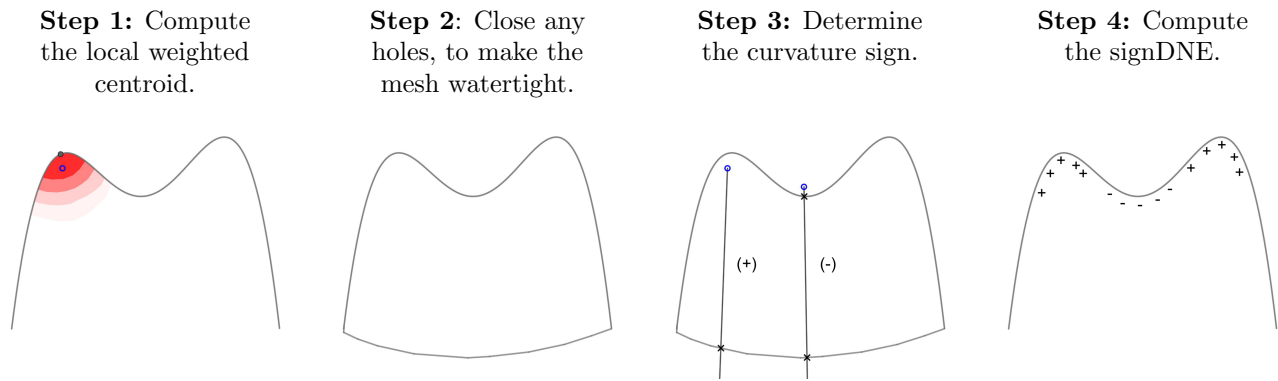


Figure \ref{fig:method} illustrates the key steps in the robust signDNE method. The intuition behind this method is based on the behavior of the centroid of the neighboring points relative to the surface concavity. When the surface bends inward at a vertex, the centroid of the neighboring poitns will likely lie outside of the surface, indicating concavity. Conversely, when the surface bends outward, the centroid of the neighboring points will likely lie inside of the surface, indicating convexity. Additionally, the algorithm for ray casting is less sensitive to changes in mesh representation.

\section{Package features and interface}
The Python package includes both the original ariaDNE and the new sign-oriented extension. Users can use the package both as a standard Python library and as a standalone command-line interface (CLI).
The core functionality is encapsulated in the function \textbf{ariaDNE}, which outputs both the original DNE value and the signDNE values.

\noindent
The \textbf{ariaDNE} function takes the following inputs:
\begin{enumerate}
	\item Mesh in the format of the Trimesh library. If the mesh is not watertight, a watertight version of the mesh is generated on the fly to use for ray casting.
	\item Optional bandwidth, that is the $\epsilon$ value for the Gaussian kernel. Default is set to be $0.08$.
	\item Optional distance cutoff for the local neighbourhoods. Default is $0$.
	\item Optional desired distance type, either euclidean or geodesic. Default is Euclidean.
	\item Optional pre-computed distances. The format of which should be a symmetric $n\times n$  matrix with pairwise distances, where $n$ is the number of points. 
\end{enumerate}

\noindent
The \textbf{ariaDNE} function returns the following values:
\begin{enumerate}
	\item local\_curvature,  which is an ordered list of the signed local bending estimates for each vertex.
	\item local\_dne, which is local\_curvature weighted by the vertex area. The vertex area is defined as the average area of the adjacent triangular faces. 
	\item dne, which is the original DNE value. This is also the sum of local\_dne.
	\item positive\_dne, which is the positive component of DNE.
	\item negative\_dne, which is the negative component of DNE.
	\item surface\_area, which is the total surface area of the input mesh.
	\item positive\_surface\_area, which is the surface area of the positive DNE regions.
	\item negatives\_surface\_area, which is the surface area of the negative DNE regions.
\end{enumerate}

The CLI offers a convenient interface for directly processing files using this function, streamlining workflows and enhancing usability for different applications.
The CLI features batch processing, namely, recursive folder processing and multiple file processing. By default, results are written to STDOUT, but can optionally be exported as a CSV file. Detailed documentation of how to use the CLI is also provided in the code repository. Below are examples demonstrating basic CLI features.

Calculating DNE or signDNE values for multiple files and save result as a file:
\begin{verbatim}
	signDNE path/to/mesh1.obj path/to/mesh2.ply -o DNEs.csv
\end{verbatim}

Calculate DNE or signDNE for all files in a folder with a custom bandwidth:
\begin{verbatim}
	signDNE path/to/mesh/directory -b 0.1
\end{verbatim}

Visualization is also featured for single-file inputs. The mesh is color-graded based on normalized local curvature values, with the following color scheme: red represents positive values, and blue indicates negative values. The intensity of the color reflects the magnitude of the local bending estimates.

An example demonstrating how to use the visualization feature is:
\begin{verbatim}
	signDNE path/to/mesh.ply -v
\end{verbatim}

In Figure \ref{fig:demovis}, we demonstrate visualization on various biological specimens. This in turn indicates the potential use of signDNE to a diverse kinds of shape data. 

\section{Test of robustness}

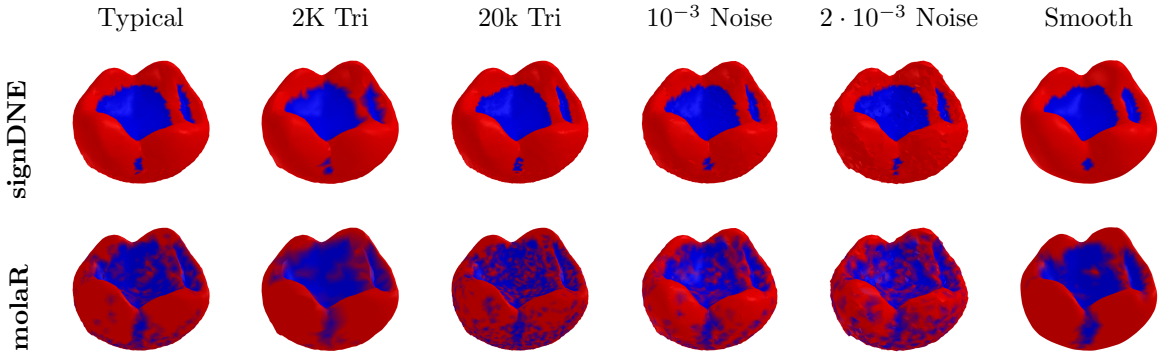
\begin{figure}[h]
	\centering
	\begin{tikzpicture}
		\matrix[matrix of nodes,
		column sep=0.5em,
		nodes={align=center}] (m)
		{
			\node {}; & 
			\node {Typical}; & 
			\node {2K Tri}; &
			\node {20k Tri}; &
			\node {$10^{-3}$ Noise}; &
			\node {$2 \cdot 10^{-3}$ Noise}; &
			\node {Smooth}; \\
			\node[rotate=90, anchor=center, xshift=0.7cm] {\textbf{signDNE}}; &
			\node {\includegraphics[width=.12\textwidth]{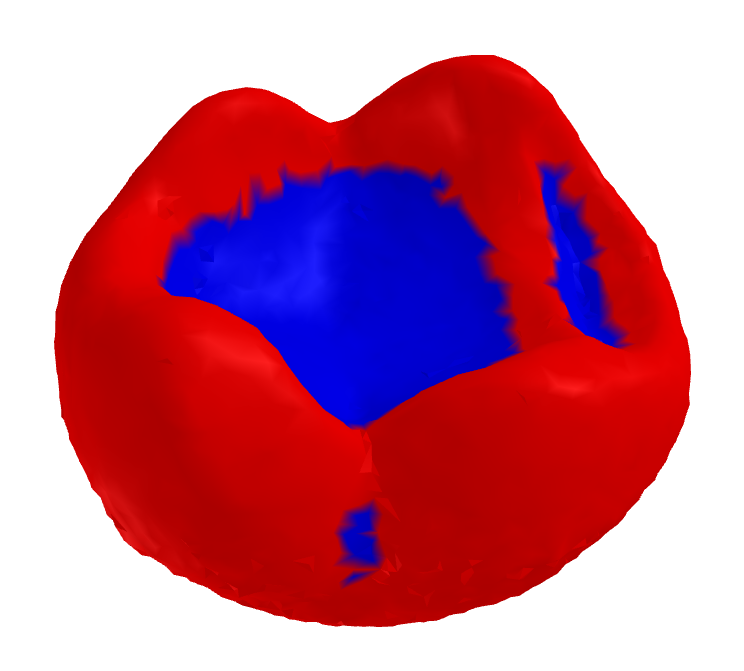}}; &
			\node {\includegraphics[width=.12\textwidth]{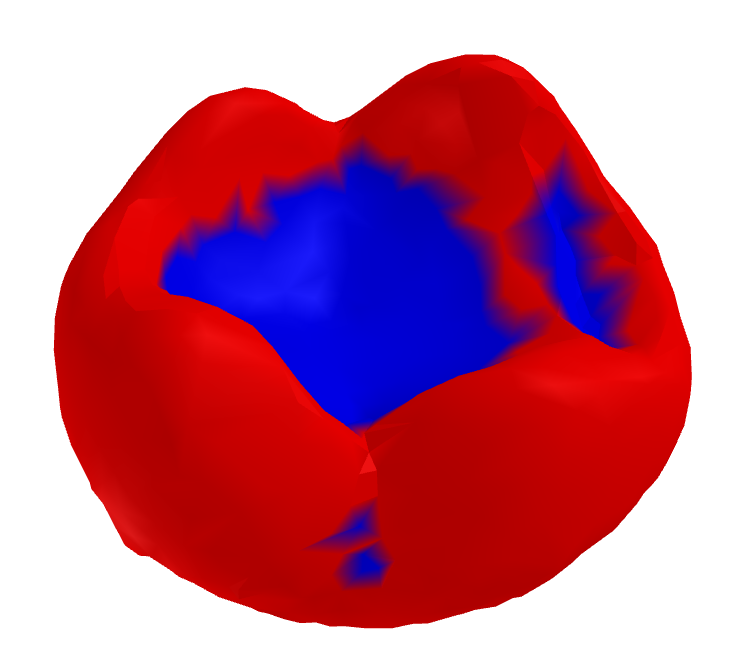}}; &
			\node {\includegraphics[width=.12\textwidth]{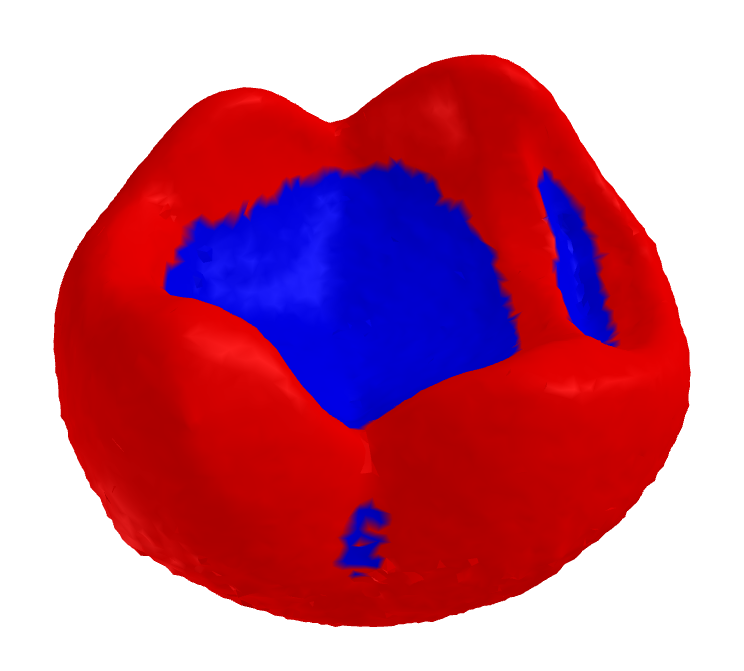}}; &
			\node {\includegraphics[width=.12\textwidth]{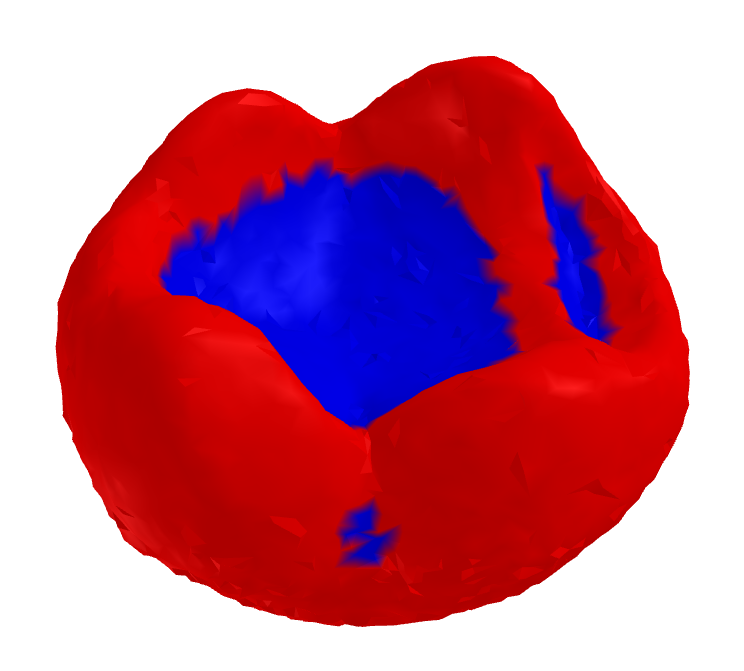}}; &
			\node {\includegraphics[width=.12\textwidth]{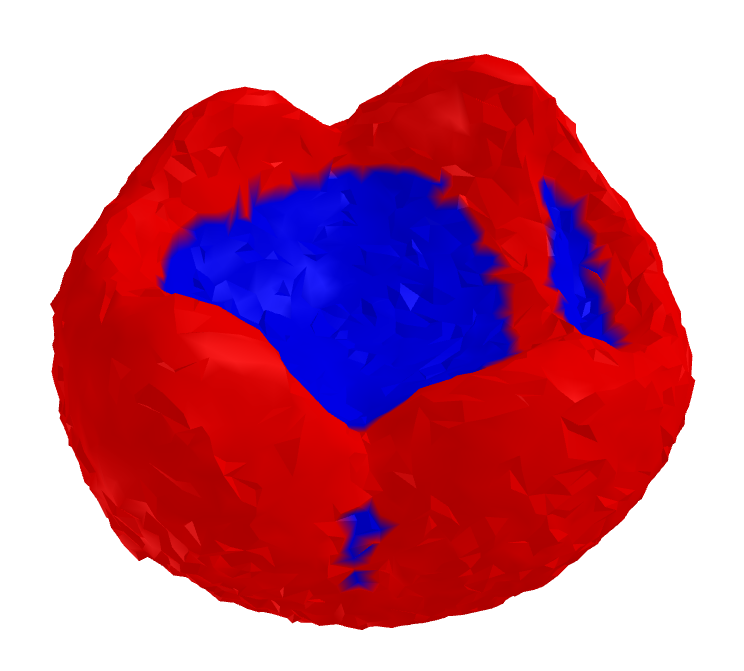}}; &
			\node {\includegraphics[width=.12\textwidth]{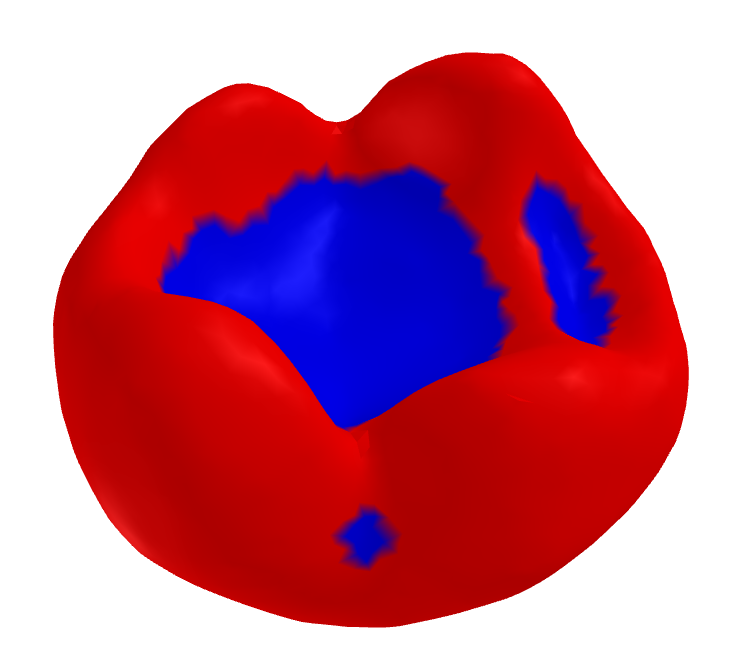}}; \\
			\node[rotate=90, anchor=center, xshift=0.7cm] {\textbf{molaR}}; &
			\node {\includegraphics[width=.12\textwidth]{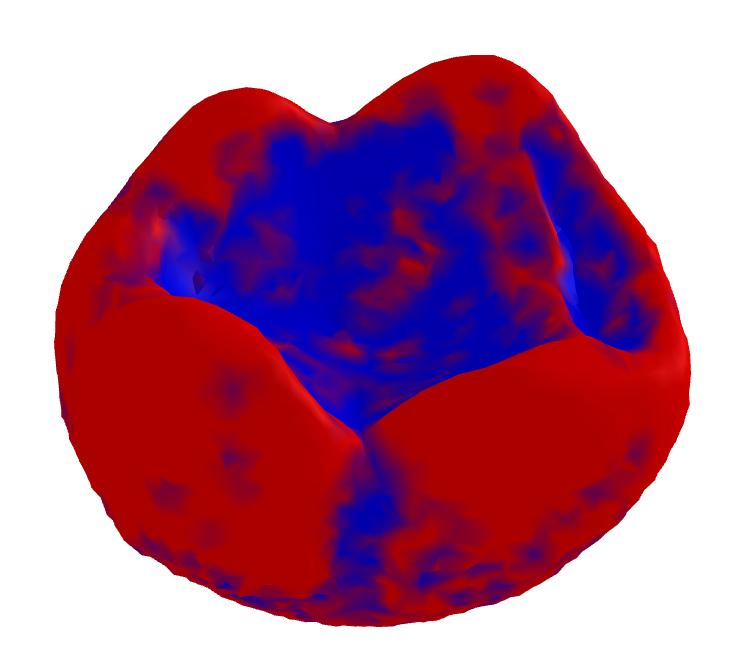}}; &
			\node {\includegraphics[width=.12\textwidth]{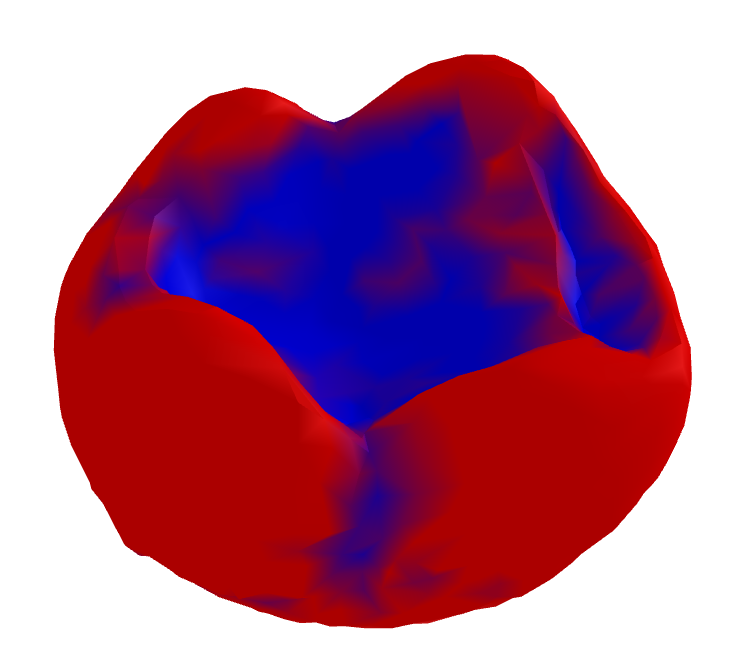}}; &
			\node {\includegraphics[width=.12\textwidth]{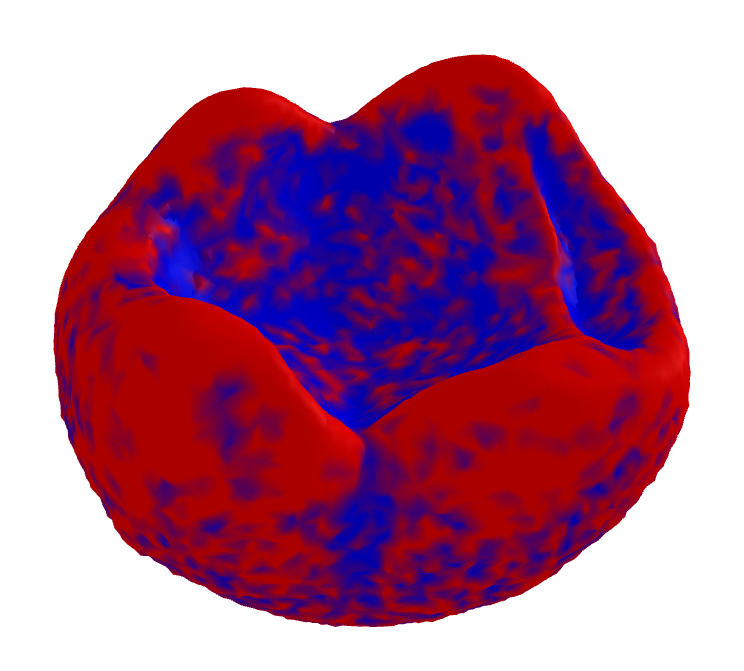}}; &
			\node {\includegraphics[width=.12\textwidth]{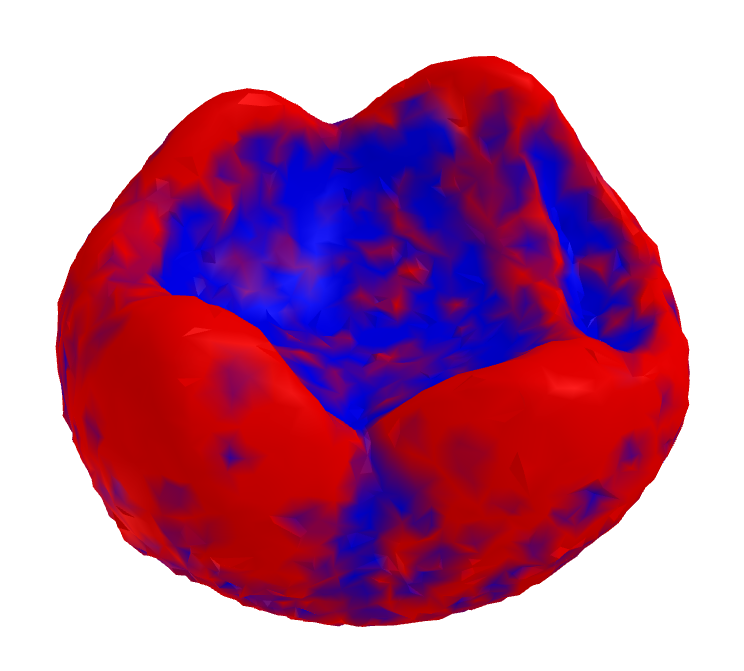}}; &
			\node {\includegraphics[width=.12\textwidth]{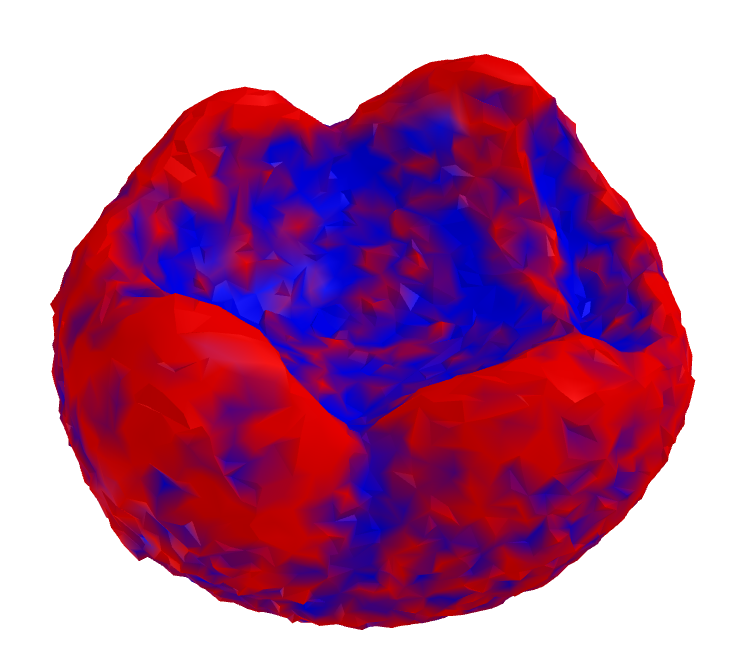}}; &
			\node {\includegraphics[width=.12\textwidth]{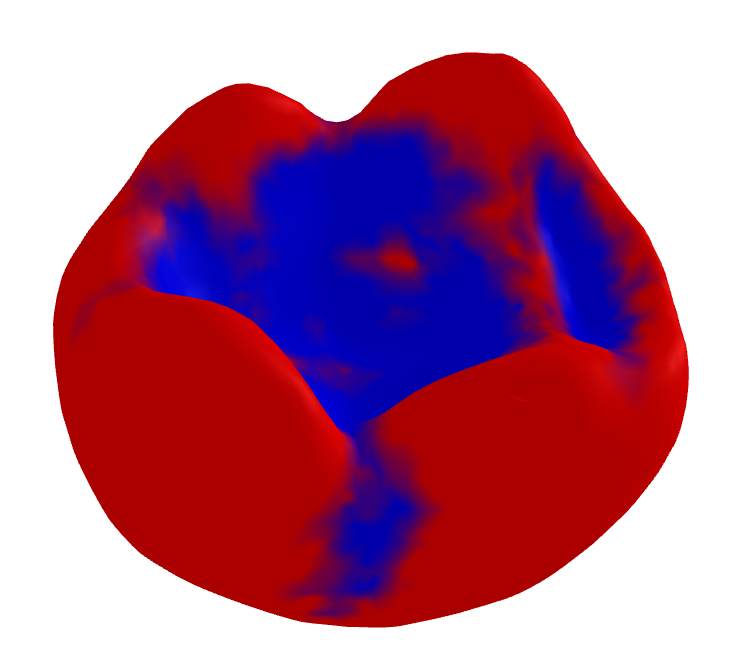}}; \\
		};
	\end{tikzpicture}
	\label{fig:comparison_visual2}
	\caption{Comparing the sign computation using our method and molaR on varying triangle count, added noise and smoothing.
	}
\end{figure}

To demonstrate the robustness of our method against varying data representations, Figures \ref{fig:comparison_visual2} and 5 compare positive and negative DNE values computed using our method with those obtained from molaR. The input data includes several perturbations on a triangular mesh, representing the same continuous surface. The perturbations are listed as follows: a mesh with approximately 2,000 triangular faces, a mesh with approximately 20,000 triangular faces, a mesh with $10^{-3}$ noise (where random values sampled from a normal distribution with mean 0 and standard deviation 0.001 are added to each coordinate), a mesh with  $2 \cdot 10^{-3}$ noise (using the same method but with a higher noise level), and a smoothed mesh. The DNE estimates are normalized according to the DNE estimate of the basis mesh, meaning we divide the DNE estimates of each perturbation by the DNE estimate of the basis mesh. 
The figures illustrate that our method exhibits greater resilience to these perturbations compared to molaR, highlighting its robustness and reliability across different data representations.

\begin{figure}[h]
	\centering
	\begin{tikzpicture}
		\matrix
		{
			\node {\includegraphics[width=0.7\linewidth]{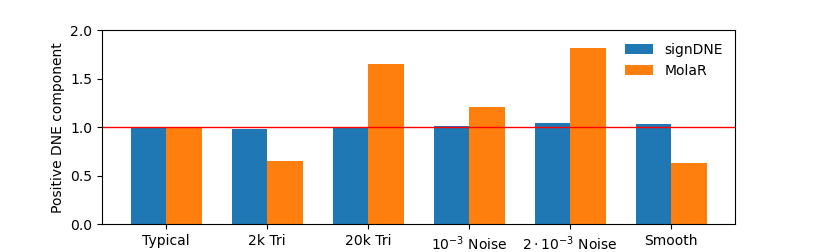}};\\
			\node {\includegraphics[width=0.7\linewidth]{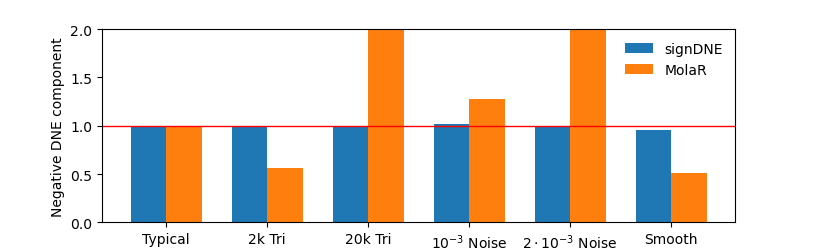}};\\
		};
	\end{tikzpicture}
	\label{fig:comparison_histo}
	\caption{Comparing signDNE computed by our method and molaR on varying triangle count, added noise and smoothing. The DNE estimates are normalized according to the DNE estimate on the basis mesh.}
\end{figure}

\begin{table}[h!]
	\centering
	\begin{tabular}{lllllll}
		\toprule
		\multicolumn{1}{c}{\textbf{Mesh}} &
		\multicolumn{2}{c}{\textbf{DNE}} &
		\multicolumn{2}{c}{\textbf{Positive DNE}} &
		\multicolumn{2}{c}{\textbf{Negative DNE}} \\
		\multicolumn{1}{c}{} &
		\multicolumn{1}{c}{\textit{Python}} &
		\multicolumn{1}{c}{\textit{MATLAB}} &
		\multicolumn{1}{c}{\textit{Python}} &
		\multicolumn{1}{c}{\textit{MATLAB}} &
		\multicolumn{1}{c}{\textit{Python}} &
		\multicolumn{1}{c}{\textit{MATLAB}} \\
		\midrule
		Typical & 0.1190 & 0.1190 & 0.1050 & 0.1050 & -0.0140 & -0.0140 \\
		2k Tri  & 0.1164 & 0.1164 & 0.1026 & 0.1026 & -0.0138 & -0.0138 \\
		20k Tri & 0.1192 & 0.1192 & 0.1054 & 0.1054 & -0.0138 & -0.0138 \\
		Noise1  & 0.1202 & 0.1202 & 0.1060 & 0.1060 & -0.0142 & -0.0142 \\
		Noise2  & 0.1230 & 0.1230 & 0.1092 & 0.1092 & -0.0138 & -0.0138 \\
		Smooth  & 0.1215 & 0.1215 & 0.1081 & 0.1081 & -0.0134 & -0.0134 \\
		\bottomrule
	\end{tabular}
	\caption{Python and MATLAB implementation comparison}
	\label{tab:compare}
\end{table}

In addition to developing a Python implementation, we have also enhanced the original ariaDNE MATLAB package by integrating the new sign-oriented DNE feature. To ensure consistency between the two implementations, we experimented with comparing the output values of both the Python package and the MATLAB version. Table \ref{tab:compare} presents the computed values for the previous dataset with five different perturbations of a standard mesh. The results demonstrate that the outputs from both the Python and MATLAB implementations are in complete agreement, confirming the accuracy and consistency of our cross-platform solution.


\section{The bandwidth parameter}
\begin{figure}
	\centering
	\begin{tikzpicture}
		\matrix[matrix of nodes,
		column sep=1em,
		nodes={align=center}] (m)
		{
			\node {}; & 
			\node {.04}; &
			\node {.06}; &
			\node {.08}; &
			\node {.10}; &
			\node {.12}; &
			\node {.14}; \\
			\node[rotate=90, anchor=center, xshift=0.7cm] {\textbf{signDNE}}; &
			\node {\includegraphics[width=.13\textwidth]{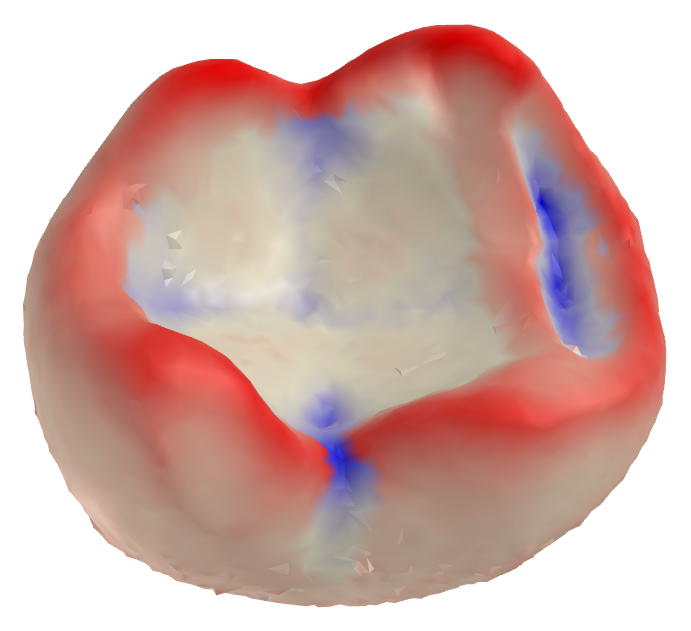}}; &
			\node {\includegraphics[width=.13\textwidth]{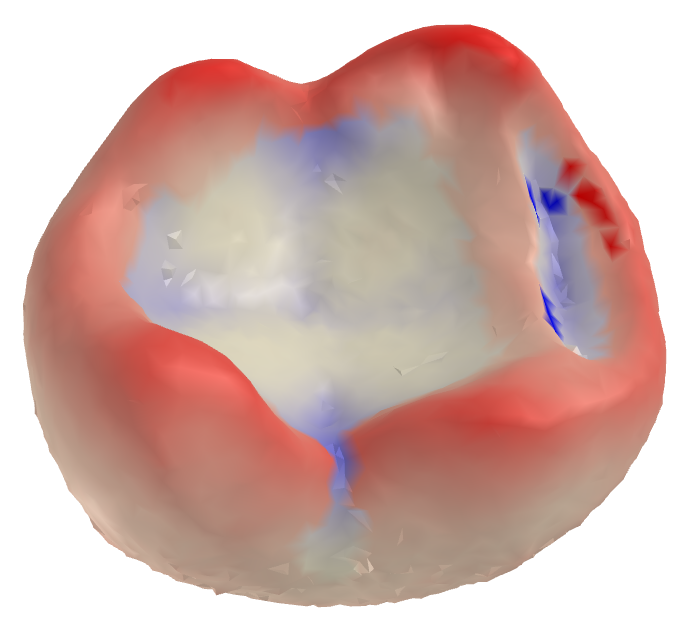}}; &
			\node {\includegraphics[width=.13\textwidth]{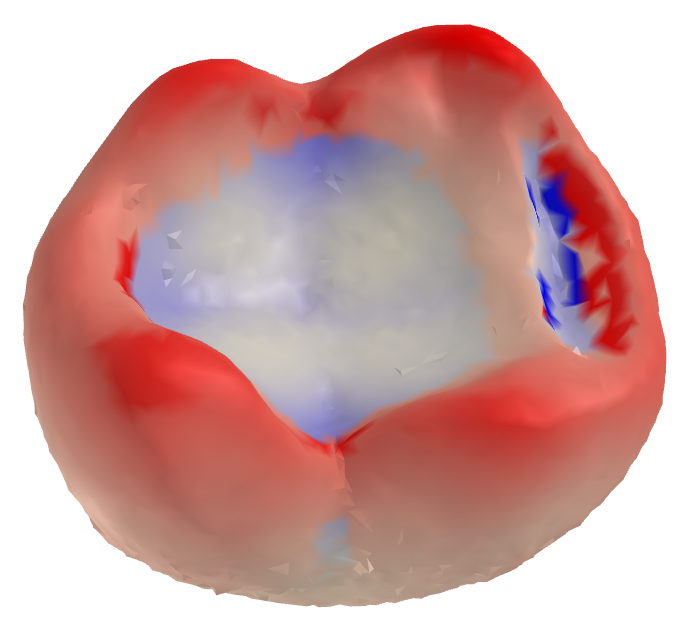}}; &
			\node {\includegraphics[width=.13\textwidth]{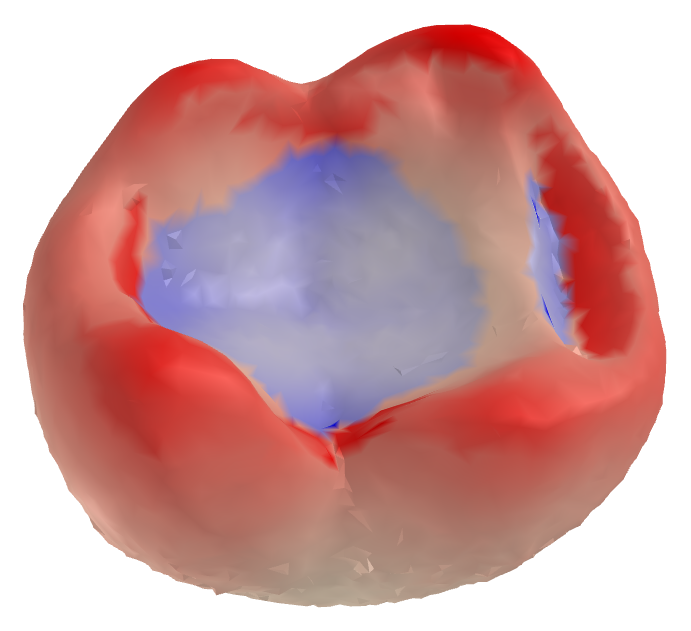}}; &
			\node {\includegraphics[width=.13\textwidth]{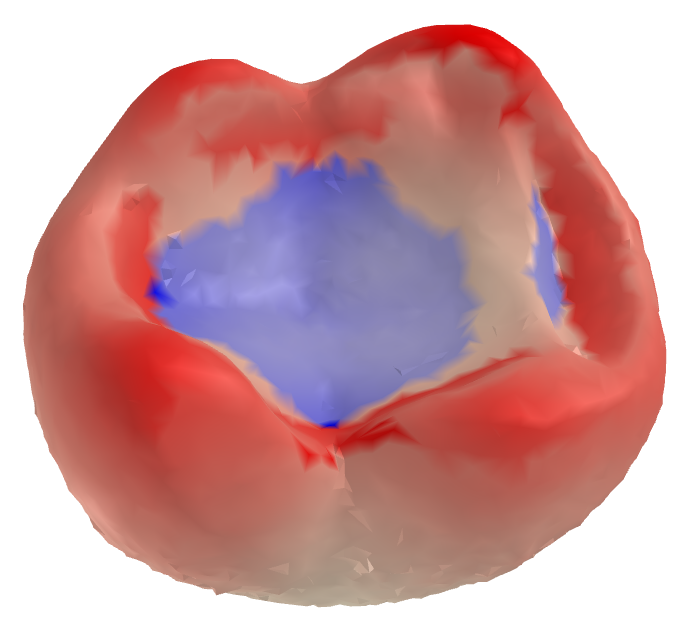}}; &
			\node {\includegraphics[width=.13\textwidth]{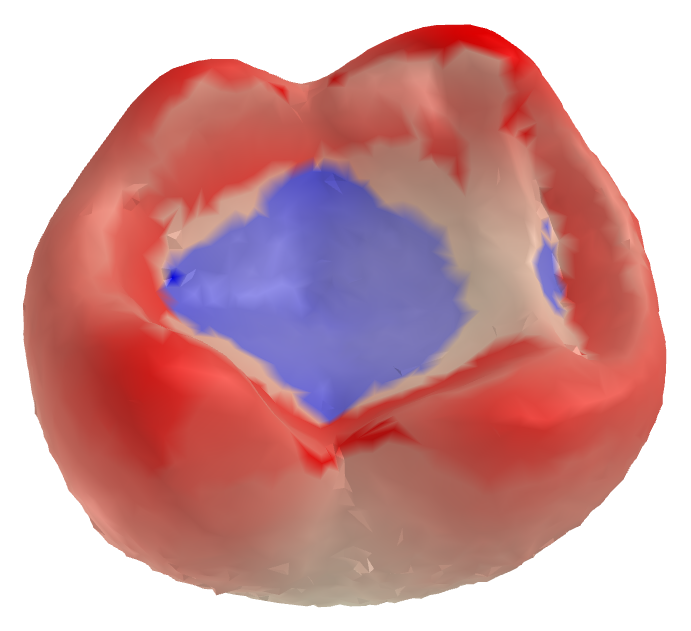}}; \\
			\node[rotate=90, anchor=center, xshift=0.7cm] {\textbf{Gaussian}}; &
			\node {\includegraphics[width=.125\textwidth]{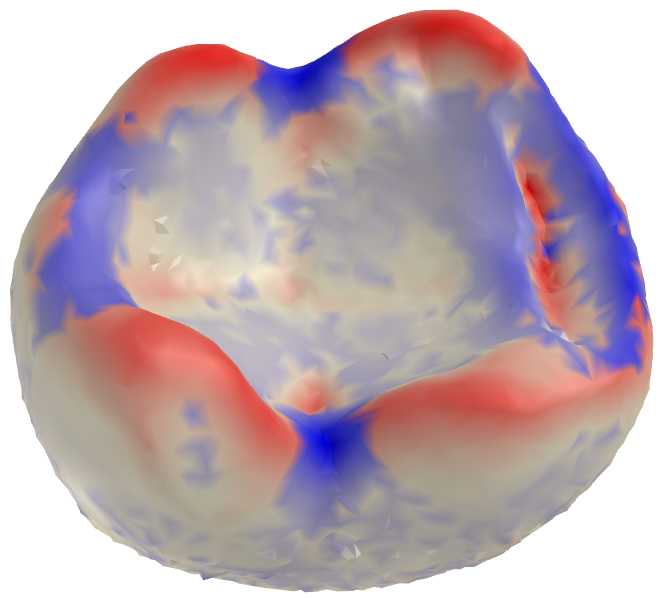}}; &
			\node {\includegraphics[width=.125\textwidth]{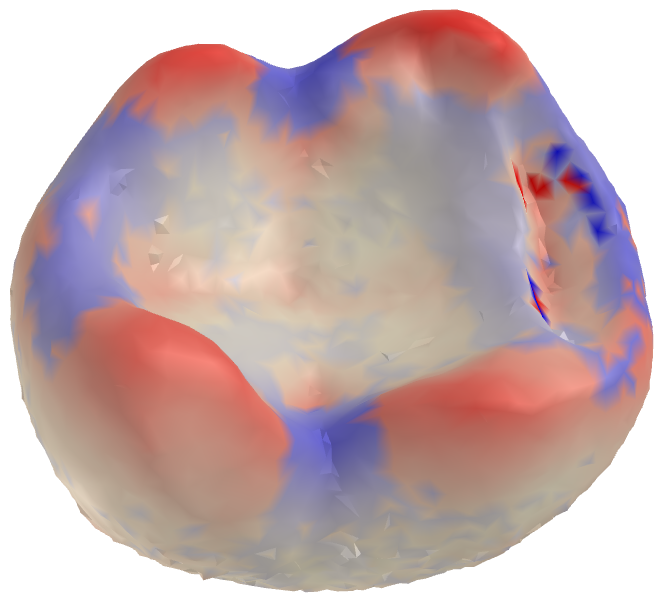}}; &
			\node {\includegraphics[width=.125\textwidth]{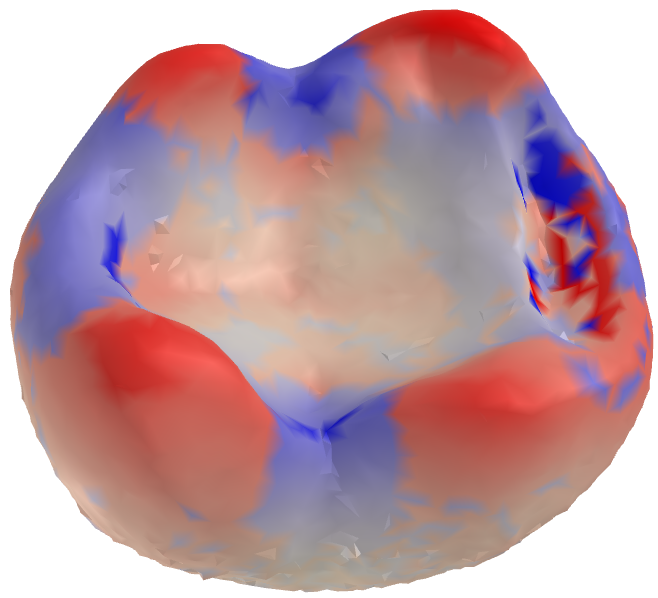}}; &
			\node {\includegraphics[width=.125\textwidth]{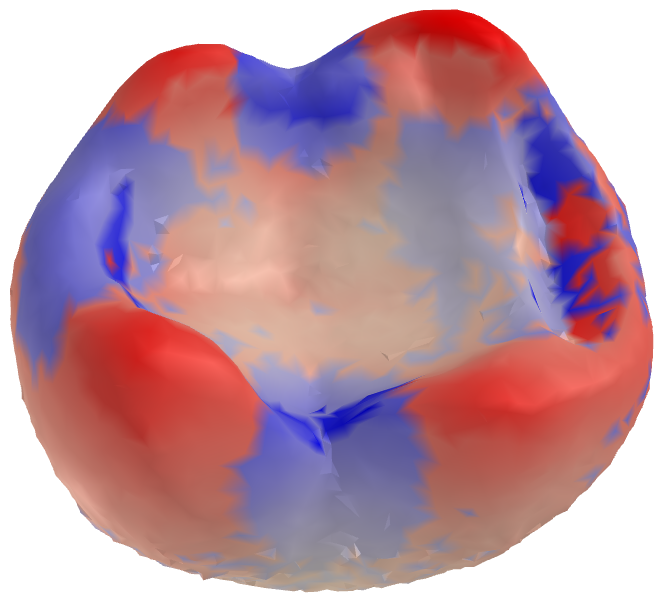}}; &
			\node {\includegraphics[width=.125\textwidth]{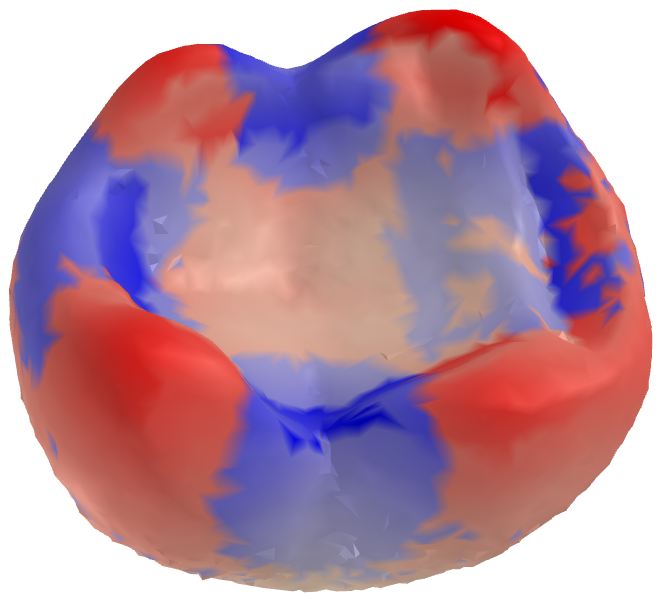}}; &
			\node {\includegraphics[width=.125\textwidth]{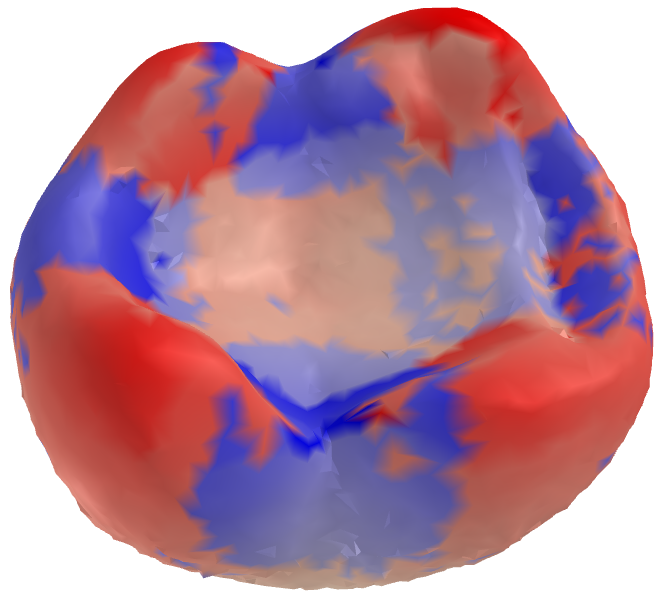}};
			\\
			\node[rotate=90, anchor=center, xshift=0.7cm] {\textbf{Mean}}; &
			\node {\includegraphics[width=.13\textwidth]{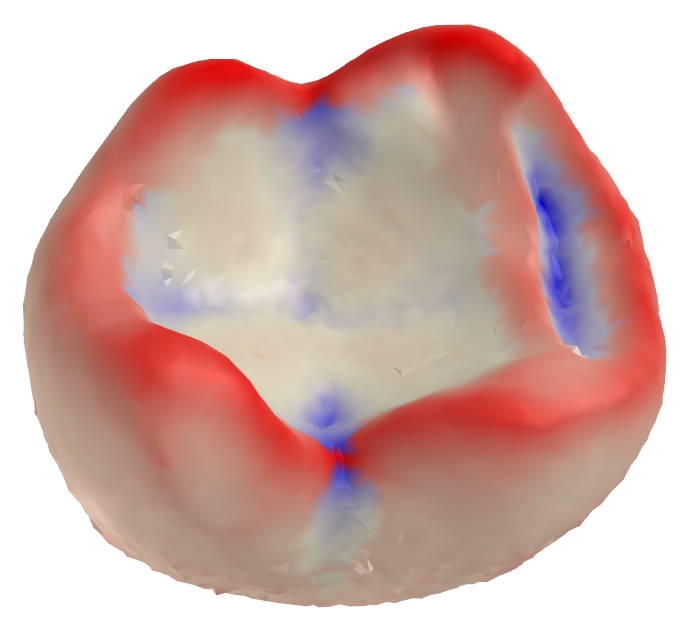}}; &
			\node {\includegraphics[width=.13\textwidth]{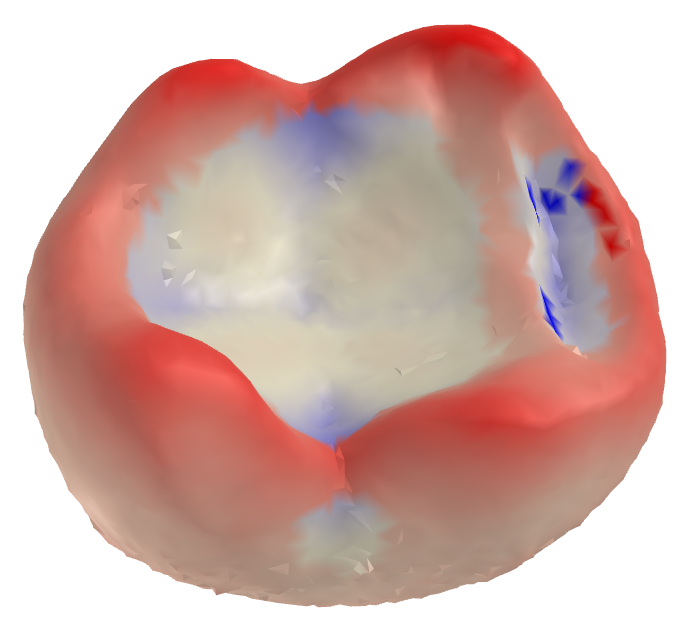}}; &
			\node {\includegraphics[width=.13\textwidth]{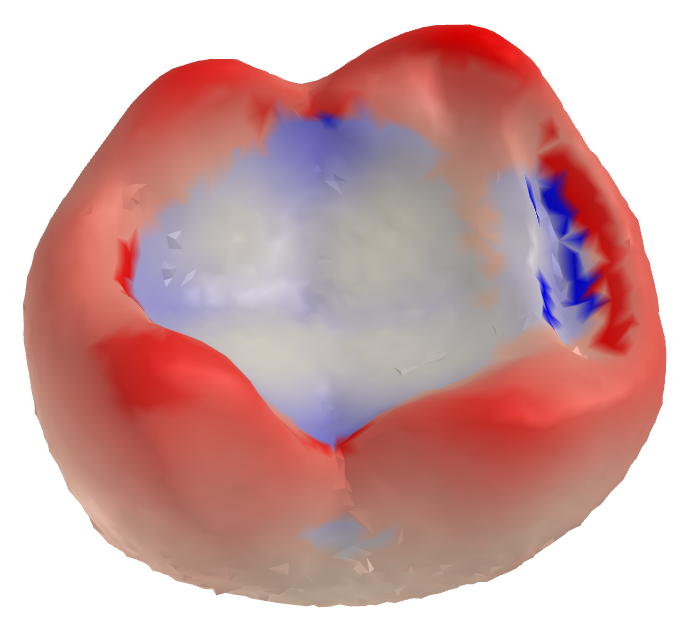}}; &
			\node {\includegraphics[width=.13\textwidth]{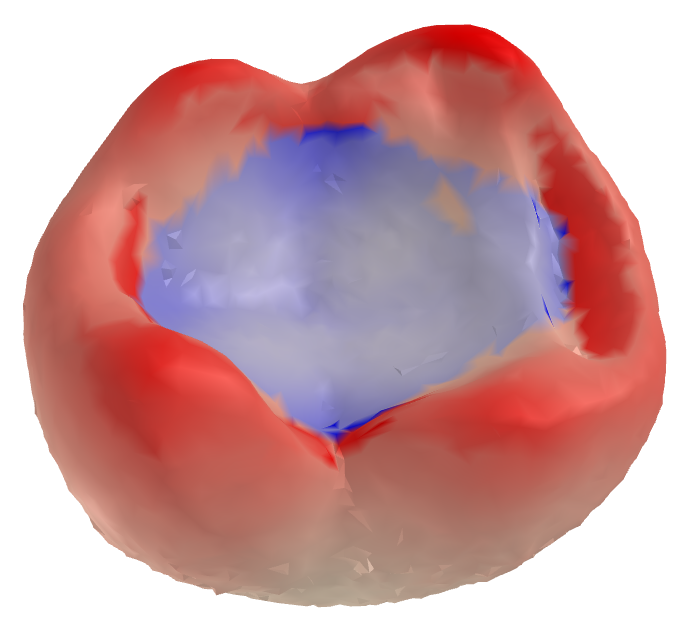}}; &
			\node {\includegraphics[width=.13\textwidth]{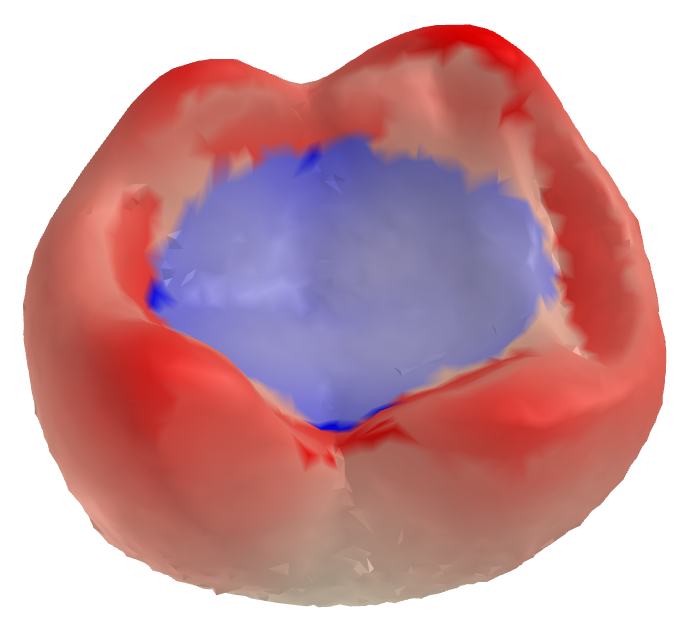}}; &
			\node {\includegraphics[width=.13\textwidth]{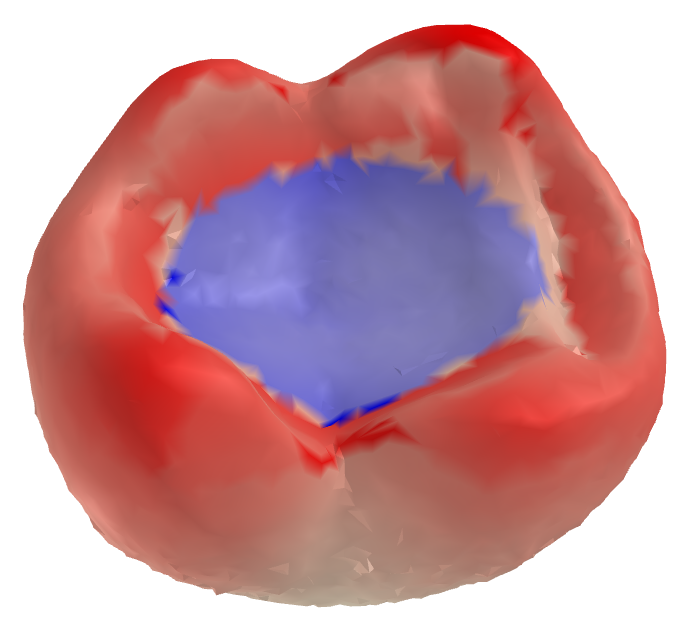}};
			\\
		};
	\end{tikzpicture}
	\label{fig:compare_sign}
	\caption{Visualization comparing three methods. In the top row, the signs are calculated by our method. In the middle row, the signs are calculated using Gaussian mean curvature. In the bottom row, the signs are calculated using discrete mean curvature. The figure shows varying bandwidth and radius, respectively.}
\end{figure}
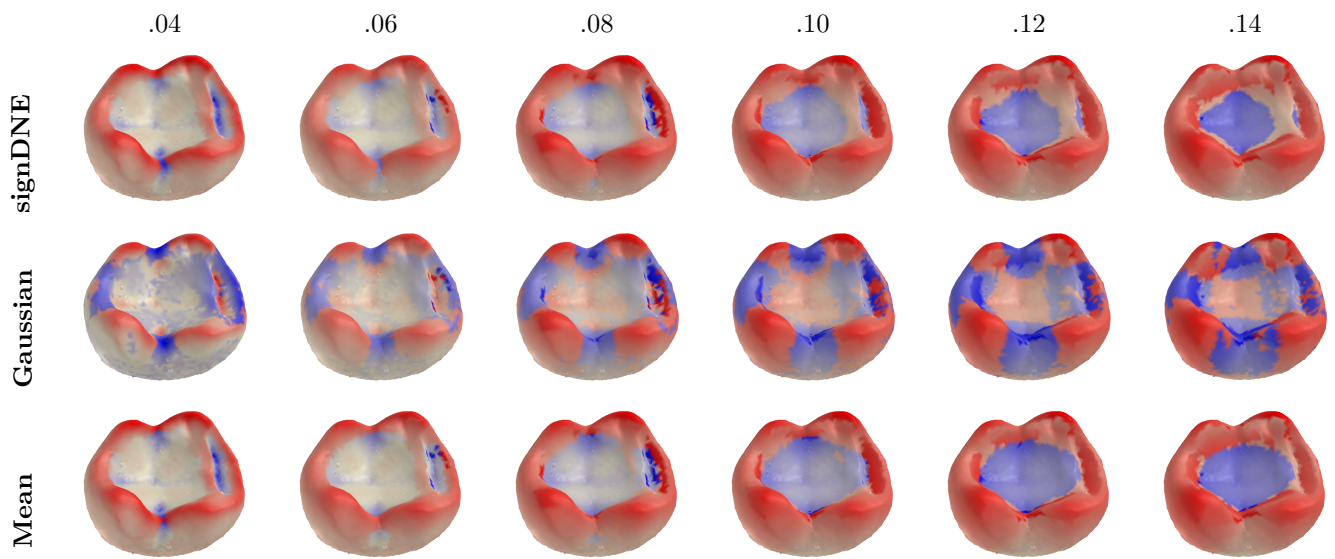

As in the original ariaDNE, users can adjust the bandwidth parameter to control the scale of features they wish to analyze. A larger bandwidth captures features of a greater scale, such as cusps or valleys, while a smaller bandwidth focuses on finer surface details. This flexibility in choosing bandwidth expands the scope of potential research questions, allowing for more in-depth analysis beyond the original focus on dental topography. In Figure \ref{fig:compare_sign}, we illustrate the effects of varying the bandwidth parameter on the local curvature field over the surface. As expected, a larger bandwidth results in higher local DNE values (more intense color gradients), emphasizing larger-scale features. Further, a smaller bandwidth results in lower local DNE values, emphasizing finer-scale features. 

In Figure 6, we compare our method with conventional methods like discrete Gaussian curvature (middle row) and discrete mean curvature (bottom row), using their standard implementation from the Trimesh Python package based on (\cite{cohen2003restricted}). Gaussian curvature, by definition, assigns negative values to saddle points and positive values to cusps. However, as the radius parameter increases (analogous to ariaDNE's bandwidth), it inconsistently categorizes the saddle point regions, failing to accurately capture key biological features. While discrete mean curvature performs better, it still misclassifies some areas, such as incorrectly assigning positive values throughout the upper tooth groove at larger radii. In contrast, our method consistently identifies the relevant biological features, demonstrating its robustness and accuracy across varying bandwidths.

\section{Discussion}

We have developed a robust Python implementation for computing DNE and its sign-oriented extension. Our approach introduces a novel method of assigning signs that reflect the concavity or convexity of the local DNE field across a surface. This method, compared with conventional approaches such as MolaR, exhibits greater resilience to variations in discrete data representation, including differences in mesh resolution, surface smoothness, and small amounts of noise. The new Python implementation includes enhanced visualization capabilities, which we have applied to a diverse range of biological shapes to showcase its potential applications. We believe this sign-oriented version could prove highly valuable in modeling and understanding ecological and functional signals in evolutionary biology.

In particular, in tasks where shape characterization is necessary to answer questions or test hypotheses, being able to distinguish between DNE from concave and convex regions is often critical. To give the prime example from the tooth literature, it is typically hypothesized that DNE values are associated with differences in dietary preference because higher DNE values reflect sharper teeth that are better for puncturing and slicing, while lower values reflect flatter teeth that are better for crushing and grinding.  Thus, higher values of the cumulative unsigned DNE value for a tooth can be reflective either of sharper and bigger points and ridges, deeper and more constricted valleys, or a combination of both.  However, it is only the ridges and points that can contribute functionally to puncturing and shearing. Therefore, it is possible that concave features can either create spurious correlations or add noise (\cite{pampush2022sign}) relative to the desired proxy. Pampush et al. (2022) proposed that uniquely high DNE values of hominid teeth were not reflective of a unique diet, but due to extensive concave areas via many small-scale crenulations of enamel on the tooth surface. They showed that when breaking down the DNE into concave and convex components, hominid teeth had more typical values of convex DNE and very high values of concave DNE. Looking at just the convex DNE allowed hominids to be more correctly classified to diet categories through comparison to non-hominid primates.  In this example, the implication is that only the convex DNE matters for the demands of diet on tooth function, and that concave DNE can be discarded. However, it's easy to imagine other hypotheses where it is instead of the concave DNE that is of interest. Knowing the relative areas of convex and concave regions can also provide important shape information. Independent of curvature magnitudes, different relative areas of convex versus concave DNE may imply differently shaped teeth that function in different ways. For instance, a tooth with equal areas of convex and concave regions is probably more versatile in the functional demands it can handle compared to one that has much more convex area or one that has much more concave area.

We note that while total DNE is always scaled to a unit area of 1 and thus comparable among and between samples of objects processed by our software package, this is not the case for convex and concave DNE calculations. Instead these typically reflect different, and complementary proportions of the unit surface (eg. 20\%-80\%, 30\%-70\%, 40\%-60\%). This means that a similar overall convex DNE calculation could be attained through either (1) limited regions with strong convexity on an otherwise concave tooth, or (2) large regions of weak convexity on a tooth with limited concave areas. Therefore comparisons of convex DNE between teeth or comparisons of convex to concave DNE in the same tooth, should control for the area of DNE. We chose not to impose a control for this in our software output in the interest of greater transparency and to accommodate more potential use cases. As calculated, the sum of the convex and concave DNE equals the total DNE. If we had controlled for the area in our signed calculations, this would no longer be true. Furthermore, depending on the question, researchers may opt for different ways of controlling or not to control.

\section*{Data/code availability}
The Python package signDNE and the data used in the manuscript are available from the GitHub repository \url{https://github.com/frisbro303/signDNE_python}. 
The updated version of the MATLAB implementation of ariaDNE from the GitHub repository \url{https://github.com/frisbro303/signDNE_MATLAB}. Both are available free to download and use.

\section*{Acknowledgements}
DB was partially supported by NSF BCS 1552848 and NSF DBI 1759839. ID acknowledges the support of the Math+X grant 400837 from the Simons Foundation. 

\bibliographystyle{unsrt}
\bibliography{references.bib}
\end{document}